\documentclass{article}
\usepackage[dvips]{graphicx}
\begin{document}
\title{Dynamics of positive- and negative-mass solitons in optical lattices
and inverted traps}
\author{H Sakaguchi$^{\dag}$ and B A Malomed$^{\ddag}$,\\
$^{\dag}$Department of Applied Science for Electronics and Materials,\\
Interdisciplinary Graduate School of Engineering Sciences,\\
Kyushu University, Kasuga, Fukuoka 816-8580, Japan\\
$^{\ddag}$Department of Interdisciplinary Studies, Faculty of Engineering,\\
Tel Aviv University, Tel Aviv 69978, Israel}
\maketitle
\begin{center}
{\Large Abstract}
\end{center}
We study the dynamics of one-dimensional solitons in the attractive and
repulsive Bose-Einstein condensates (BECs) loaded into an optical lattice
(OL), which is combined with an external parabolic potential. First, we
demonstrate analytically that, in the repulsive BEC, where the soliton is of
the gap type, its effective mass is \emph{negative}. This gives rise to a
prediction for the experiment: such a soliton cannot be not held by the
usual parabolic trap, but it can be captured (performing harmonic
oscillations) by an \emph{anti-trapping} inverted parabolic potential. We
also study the motion of the soliton a in long system, concluding that, in
the cases of both the positive and negative mass, it moves freely, provided
that its amplitude is below a certain critical value; above it, the
soliton's velocity decreases due to the interaction with the OL. At a late
stage, the damped motion becomes chaotic. We also investigate the evolution
of a two-soliton pulse in the attractive model. The pulse generates a
persistent breather, if its amplitude is not too large; otherwise, fusion
into a single fundamental soliton takes place. Collisions between two
solitons captured in the parabolic trap or anti-trap are considered too.
Depending on their amplitudes and phase difference, the solitons either
perform stable oscillations, colliding indefinitely many times, or merge
into a single soliton. Effects reported in this work for BECs can also be
formulated for optical solitons in nonlinear photonic crystals. In
particular, the capture of the negative-mass soliton in the anti-trap
implies that a bright optical soliton in a self-defocusing medium with a
periodic structure of the refractive index may be stable in an 
\emph{anti-waveguide}.

\section{Introduction}

An effective periodic potential in the form of an optical lattice (OL),
created as an interference pattern between laser beams illuminating a
Bose-Einstein condensate (BEC), is a powerful tool that facilitates
experimental studies of many dynamical properties of BECs. Among nonlinear
dynamical perturbations that BECs may support are bright solitons, that were
recently created in the experiment \cite{soliton_exper}. Direct observation
of solitons in a BEC loaded in the OL has not yet been reported, although
spitting of wave packets in a $^{87}$Rb condensate, confined in a
cylindrical trap with a weak longitudinal OL potential, which was observed
in a very recent work \cite{soliton_transient}, can be explained as a
manifestation of transient soliton dynamics. In anticipation of progress in
the experiments, a lot of theoretical work has been done on the topic of BEC
solitons in OLs (\textquotedblleft solitons\textquotedblright\ are here
realized as robust solitary waves, rather than as exact solutions of
integrable equations). In particular, the use of OLs opens a way to observe 
\textit{gap solitons}, that are expected to exist when the interaction
between atoms in the BEC is repulsive (hence ordinary bright solitons are
not possible) \cite{SalernoJPhysB,othergapsol}. Moreover, while only
one-dimensional (1D) solitons have been thus far observed in BECs, it was
demonstrated that OLs may support multidimensional solitons of the gap type
in the case of repulsion \cite{SalernoJPhysB}, and, which is plausibly more
relevant to the experiment, they can also stabilize multidimensional 2D and
even 3D solitons in an \emph{attractive} BEC, which formally exist without
the periodic potential, but are unstable because of the collapse \cite{BMS}.
Besides that, bright solitons with an internal \emph{vorticity} can be
stabilized too by the OL in the 2D case \cite{BMS,YangMuss}. OLs can also
affect solitons in BECs in various other ways \cite{otherOLsol,Tashkent}.

Similar dynamical features are expected from optical solitons in nonlinear
photonic crystals (NPCs) \cite{PhotCryst,BMS,YangMuss}. In the latter case,
a counterpart of the OL is periodic modulation of the local refractive index
in the transverse direction(s) \cite{Wang,BMS,YangMuss}. 2D solitons can be
easily stabilized in the NPC \cite{PhotCryst} (in the uniform nonlinear
medium, it would be the well-known Townes soliton, which is unstable due to
the collapse \cite{Berge}).

Thus far, most studies of solitons in BECs loaded in the OL did not
systematically consider their motion, nor two-soliton structures (very
recently, mobility of discrete solitons, which represent the BEC solitons in
OLs in the tight-binding approximation, was addressed, vis-a-vis the
corresponding Peierls-Nabarro potential, in Ref. \cite{discr_solitons}). We
aim to analyze these issues in the 1D case. Several results will be
reported. First, both ordinary solitons in the case of attraction, and gap
solitons in the BEC with repulsion (the latter is more common in BECs \cite
{Pit}) move across the OL freely if the soliton's amplitude (i.e., the
number of atoms in it) does not exceed a certain critical value. Above that
value, the soliton feels action of an effective braking force, induced by
the Peierls-Nabarro potential, and eventually comes to a halt. It is
relevant to mention that a similar property is known for moving lattice
solitons in the discrete nonlinear Schr\"{o}dinger (NLS) equation and
related models: the solitons move freely unless they are too
\textquotedblleft heavy" \cite{Feddersen,Dima}. For the solitons whose
motion is hindered, we demonstrate dependence of the braking rate on the
soliton's amplitude. At a later stage of the braking, the soliton remains a
coherent object, while its motion becomes chaotic.

We also consider dynamics of a pulse which, in the absence of the OL
potential, would give rise to a well-known two-soliton (breather solution)
in the NLS equation (with attraction). We demonstrate that, if the initial
amplitude of the two-soliton is below a critical value, it indeed develops
into a persistent breather in presence of the OL. However, the two-soliton
with the amplitude exceeding the critical value relaxes into a stationary
(fundamental) soliton.

It is known that the dispersion law for linear excitations, whose wavenumber
is close to the edge of the Brillouin zone (BZ) induced by the OL potential,
is characterized by a negative effective mass (curvature of the dispersion
curve), unlike a vicinity of the BZ center, where the effective mass is
positive. This feature was recently observed in an experiment \cite
{neg_mass_experiment}, and it actually gives rise to the gap solitons in the
case of the repulsive nonlinearity \cite{SalernoJPhysB}. In this work, we
will, first of all, demonstrate that the effective mass of the \textit{
soliton proper} may also be negative. A noteworthy feature of the analysis
is that the underlying Gross-Pitaevskii (GP) equation, including the OL
potential, does not conserve momentum, while equations providing for
asymptotic description of the soliton dynamics conserve it. We provide for
an explanation of this discrepancy.

Then, we demonstrate that, having the negative mass, the gap soliton cannot
be trapped by a usual external parabolic potential; however, it readily gets
captured by an \emph{anti-trapping} (inverted parabolic) potential (a
possibility of confining a usual soliton in an inverted potential that
rapidly oscillates in time was recently proposed in Ref. \cite{Tashkent},
which, however, a completely different mechanism). This prediction suggests
an experimental verification, especially in view of the fact that ordinary
solitons in BECs were actually observed in inverted (expulsive) potentials 
\cite{soliton_exper}.

Lastly, we also consider a state in which two identical solitons (with the
positive or negative mass) oscillate in opposite directions and periodically
collide inside the, respectively, trap or anti-trap, applied on top of the
OL. We demonstrate that there is a critical value of the amplitude of the
solitons specific to this case, such that they merge into a single soliton
(after one or several collisions) if the initial amplitude exceeds the
critical value. Otherwise, the solitons keep to oscillate, colliding
indefinitely many times.

All the above-mentioned results can also be applied to optical solitons in
NPCs. In particular, in the case of the self-defocusing nonlinearity
(negative Kerr effect), the corresponding gap soliton will be stably
confined in the NPC under the anti-trapping conditions, which, in optics,
corresponds to an \textit{anti-waveguide} \cite{Gisin}. In fact, this
suggests the first possibility to create a stable bright optical soliton in
a nonlinear anti-waveguide.

The rest of the paper is organized as follows. The model and analytical
approximations for the solitons in it are presented in section 2. Numerical
results for the motion of a soliton, and for two-solitons, are collected in
section 3. Trapping of the soliton by the parabolic potential is considered
in section 4, and collisions between solitons oscillating in the trap or
anti-trap are presented in section 5. Section 6 concludes the paper.

\section{Formulation of the model and analytical approximations for solitons}

\subsection{Basic equations and momentum conservation}

The mean-field description of the BEC dynamics is based on the GP equation
for the mean-field wave function $\psi $ in three dimensions \cite{Pit},
\begin{equation}
i\hbar \frac{\partial \psi }{\partial t}=\left[ -\frac{\hbar ^{2}}{2m}\nabla
^{2}+U\left( \mathbf{r}\right) +G|\psi |^{2}\right] \psi ,  \label{GP3D}
\end{equation}
where $m$ is the atomic mass, $G\equiv 4\pi \hbar ^{2}a/m$, with $a$ being
the $s$-wave scattering length, and $U$ is the external potential. As it was
demonstrated in a number of works \cite{1D}, in the case of a quasi-1D
(\textquotedblleft cigar-shaped\textquotedblright ) configuration, Eq. (\ref
{GP3D}) can be reduced to a 1D equation. In the presence of the OL potential
(the parabolic trap will be added later), the normalized equation for a 1D
mean-field wave function $\phi (x,t)$ is \cite{SalernoJPhysB,othergapsol}: 
\begin{equation}
i\frac{\partial \phi }{\partial t}=-\frac{1}{2}\frac{\partial ^{2}\phi }{
\partial x^{2}}+\left[ \sigma |\phi |^{2}-\varepsilon \cos \left( 2\pi
x\right) \right] \phi ,  \label{GP}
\end{equation}
where $\sigma =+1$ and $-1$ for $a>0$ and $a<0$, respectively (i.e.,
repulsion and attraction, respectively), the period of the OL is set to be 
$1$, and $-\varepsilon $ is its strength. For the case of attraction,
dynamics of solitons in this model was first investigated (in terms of
nonlinear optics) in Ref. \cite{Wang}.

It is necessary to note that, while all the works published to date describe
the gap solitons within the framework of the mean-field GP equation, effects
of quantum depletion (\textquotedblleft evaporation\textquotedblright\ of
atoms from the soliton wave function due to quantum fluctuations) may affect
them similar to the way they were predicted to blur the notch of dark
solitons in BECs \cite{depletion}, as gap solitons contain many notches
[see. e.g., Fig. 1(b) below]. However, analysis of the quantum depletion is
beyond the scope of the present work.

First, we aim to give an explanation to the negativeness of the
gap-soliton's mass in the case of $\sigma =+1$ (repulsion). To this aim, we
notice that the gap soliton may be represented by a combination of two
terms, each being a product of a slowly varying amplitude $u(x,t)$ or 
$v(x,t) $ and rapidly varying carriers, $\exp \left( \pm i\pi x\right) $: 
\begin{equation}
\phi (x,t)=\left( \sqrt{\varepsilon }/2\right) \left[ U(x,t)e^{i\pi
x}+V(x,t)e^{-i\pi x}\right] .  \label{slowfast}
\end{equation}
Substituting this into Eq. (\ref{GP}), keeping only first derivatives of the
slowly varying functions, and defining rescaled variables
\begin{equation}
\tau \equiv (\varepsilon /2)t,~\xi \equiv (\varepsilon /2\pi )x
\label{tauxi}
\end{equation}
lead to the standard model (\ref{gapmodel}) that gives rise to gap solitons: 
\begin{eqnarray}
i\frac{\partial U}{\partial \tau }+i\frac{\partial U}{\partial \xi }-\sigma
\left( \frac{1}{2}|U|^{2}+|V|^{2}\right) U+V &=&0,  \nonumber \\
i\frac{\partial V}{\partial \tau }-i\frac{\partial V}{\partial \xi }-\sigma
\left( \frac{1}{2}|V|^{2}+|U|^{2}\right) V+U &=&0.  \label{gapmodel}
\end{eqnarray}

Note that both the underlying GP equation (\ref{GP}) and the asymptotic
gap-soliton model (\ref{gapmodel}) conserve the norm (number of atoms). In
the former case, it is 
\begin{equation}
N\equiv \int_{-\infty }^{+\infty }\phi ^{2}(x)dx.  \label{N}
\end{equation}
The substitution of the waveform (\ref{slowfast}) into this expression and
neglecting terms like $\int_{-\infty }^{+\infty }U(x)V^{\ast }(x)e^{2i\pi
x}dx$ (they are exponentially small, being integrals of products of rapidly
oscillating functions and slowly varying ones) yield $N=\left( \pi /2\right)
\int_{-\infty }^{+\infty }\left[ |U(\xi )|^{2}+|V(\xi )|^{2}\right] dx$,
which is indeed the norm conserved by Eqs. (\ref{gapmodel}).

On the other hand, Eqs. (\ref{gapmodel}) conserve the momentum,
\begin{equation}
P=i\int_{-\infty }^{+\infty }\left( \frac{\partial U^{\ast }}{\partial \xi}
U+\frac{\partial V^{\ast }}{\partial \xi }V\right) d\xi   \label{Pgap}
\end{equation}
(the asterisk stands for the complex conjugation), while the momentum of Eq.
(\ref{GP}), 
\begin{equation}
P_{\mathrm{GP}}=2i\int_{-\infty }^{+\infty }\frac{\partial \phi ^{\ast }}
{\partial x}\phi ~dx,  \label{Pphi}
\end{equation}
is not conserved in the presence of the periodic potential, but rather
evolves according to the equation
\begin{equation}
\frac{dP_{\mathrm{GP}}}{dt}=-4\pi \varepsilon \int_{-\infty }^{+\infty
}|\phi (x)|^{2}\sin \left( 2\pi x\right) dx,  \label{dP/dt}
\end{equation}
(integration by parts was used to simplify the expression). Making use of
Eqs. (\ref{slowfast}) and (\ref{tauxi}) and dropping exponentially small
terms of the above-mentioned type, one can cast Eq. (\ref{dP/dt}) in the form
\begin{equation}
\frac{dP_{\mathrm{GP}}}{d\tau }=2\pi ^{2}i\int_{-\infty }^{+\infty }\left(
U^{\ast }V-UV^{\ast }\right) d\xi .  \label{dP/dtau}
\end{equation}

To explain the apparent contradiction between the conservation of $P$ and
non-conservation of $P_{\mathrm{GP}}$, we note that the substitution of Eqs.
(\ref{slowfast}) and (\ref{tauxi}) into Eq. (\ref{Pphi}) yields, aside from
exponentially small terms, an expression
\[
P_{\mathrm{GP}}=\int_{-\infty }^{+\infty }\left[ i\left( \frac{\partial
U^{\ast }}{\partial \xi }U+\frac{\partial V^{\ast }}{\partial \xi }V\right)
+\pi ^{2}\left( |U|^{2}-|V|^{2}\right) \right] d\xi .
\]
Inserting this, and a relation which is a consequence of Eqs. 
(\ref{gapmodel}), 
\begin{equation}
\frac{d}{d\tau }\int_{-\infty }^{+\infty }\left( |U|^{2}-|V|^{2}\right) d\xi
=2i\int_{-\infty }^{+\infty }\left( U^{\ast }V-UV^{\ast }\right) d\xi ,
\label{U2-V2}
\end{equation}
into Eq. (\ref{dP/dt}) results in $dP/d\tau =0$.

\subsection{The mass of gap solitons}

An analytical solution to Eqs. (\ref{gapmodel}) corresponding to a gap
soliton moving at a velocity $v$, which belongs to the interval $-1<v<+1$,
is well known \cite{GapModel}. In the case of $\sigma =-1$ (the attractive
BEC), it is 
\begin{eqnarray}
U &=&\sqrt{\frac{1+v}{3-v^{2}}}\frac{2\left( 1-v^{2}\right) ^{1/4}\sin
\theta }{\sqrt{\cosh \left( 2\Xi \sin \theta \right) +\cos \theta }} 
\nonumber \\
&&\times \exp \left[ i\left( \frac{4v}{3-v^{2}}+1\right) \tan ^{-1}\left(
\tan \frac{\theta }{2}\cdot \tanh \left( \Xi \sin \theta \right) \right)
-iT\cos \theta \right] ,  \nonumber \\
V &=&-\sqrt{\frac{1-v}{3-v^{2}}}\frac{2\left( 1-v^{2}\right) ^{1/4}\sin
\theta }{\sqrt{\cosh \left( 2\Xi \sin \theta \right) +\cos \theta }} 
\nonumber \\
&&\times \exp \left[ i\left( \frac{4v}{3-v^{2}}-1\right) \tan ^{-1}\left(
\tan \frac{\theta }{2}\cdot \tanh \left( \Xi \sin \theta \right) \right)
-iT\cos \theta \right] ,  \label{uv}
\end{eqnarray}
where $\Xi \equiv \left( 1-v^{2}\right) ^{-1/2}\left( \xi -v\tau \right) $, 
$T\equiv \left( 1-v^{2}\right) ^{-1/2}\left( \tau -v\xi \right) $, and 
$\theta $, which takes values $0<\theta <\pi $, determines the amplitude and
width of the soliton. For the soliton (\ref{uv}), the momentum (\ref{Pgap})
can be calculated in an exact form: 
\begin{equation}
P_{\mathrm{sol}}=\frac{8v\sqrt{1-v^{2}}}{3-v^{2}}\left[ \frac{7-v^{2}}{
3-v^{2}}\left( \sin \theta -\theta \cos \theta \right) +\theta \cos \theta
\right] \,.  \label{Psol}
\end{equation}
The expression (\ref{Psol}) simplifies for slow solitons ($v^{2}\ll 1$),
making it possible to define the mass of the slow gap soliton as the
momentum divided by the velocity, 
\begin{equation}
m_{\mathrm{sol}}^{\mathrm{(GS)}}\equiv P_{\mathrm{sol}}/v=(8/9)\left( 7\sin
\theta -4\theta \cos \theta \right) .  \label{massGS}
\end{equation}
Note that this mass is \emph{positive} for all values of $\theta $.

In the case of the repulsive BEC, $\sigma =+1$, we obtain Eqs. (\ref
{gapmodel}) with the opposite sign in front of the cubic terms. To cast the
equations into the standard form to which the soliton solution (\ref{uv})
pertains, we apply complex conjugation to them, and define $\widetilde{U}
\equiv U^{\ast },\widetilde{V}\equiv -V^{\ast }$. Then, the gap-soliton's
momentum $\widetilde{P}$\ defined by Eq. (\ref{Pgap}) with $U$ and $V$
replaced by $\widetilde{U}$ and $\widetilde{V}$ is related to the velocity
exactly the same way as above, i.e., as per Eq. (\ref{Psol}). However, the
proper momentum $P$ is still defined by Eq. (\ref{Pgap}) in terms of the
original fields $U$ and $V$, hence $P\equiv -\widetilde{P}$. Thus, the
effective mass in the repulsive model is precisely $-m_{\mathrm{eff}}^{
\mathrm{(GS)}}$, being \emph{always negative}.

\subsection{Description in terms of the Bloch functions}

To continue the analysis, we now return to the underlying GP equation (\ref
{GP}). Another approach to solitons in the weakly nonlinear case starts with
the linear Bloch functions that obey the Mathieu equation, 
\begin{equation}
EF(x)=\left( 1/2\right) F^{\prime \prime }+\varepsilon \cos \left( 2\pi
x\right) F.  \label{F}
\end{equation}
Here $E$ is the chemical potential, and the solution is quasi-periodic,
i.e., $F(x)=e^{ikx}f(x)$, where $k$ is a wavenumber, and the function $f(x)$
is periodic, $f(x+1)=f(x)$. The solution determines the corresponding band
structure, $E=E(k)$, the points $k=0$ and $k=\pi $ being, respectively, the
center and edge of the first Brillouin zone (BZ).

Then, an approximate solution to the underlying GP equation (\ref{GP}) for a
small-amplitude broad soliton is looked for as 
\begin{equation}
\phi (x,t)=e^{-iEt}F(x)\Phi (x,t),  \label{phi}
\end{equation}
with a slowly varying envelope function $\Phi (x)$, cf. Eq. 
(\ref{slowfast}). As the norm of the solution 
(the number of atoms in the BEC), $N\equiv
\int_{-\infty }^{\infty }|\phi |^{2}dx$, approaches zero, the ordinary
soliton solution in the attractive model and its gap-soliton counterpart in
the repulsive one approach, respectively, the linear Bloch function with 
$k=0 $ and $k=\pi $.

In the general case, the slowly varying amplitude obeys an asymptotic NLS
equation, that can be derived by substituting the ansatz (\ref{phi}) into
Eq. (\ref{GP}) \cite{SalernoJPhysB}: 
\begin{equation}
i\frac{\partial \Phi }{\partial t}=-\frac{1}{2}m_{\mathrm{eff}}^{-1}\frac{
\partial ^{2}\Phi }{\partial x^{2}}+\sigma g|\Phi |^{2}\Phi .  \label{asympt}
\end{equation}
Here, the effective mass $m_{\mathrm{eff}}$ for linear excitations is
determined from the curvature of the band structure of the linear Bloch
states as $m_{\mathrm{eff}}^{-1}=E^{\prime \prime }(k)$, and $g\equiv
\int_{0}^{1}|F(x)|^{4}dx/\int_{0}^{1}|F(x)|^{2}dx$ (recall $1$ is the
normalized period of the OL potential). Notice that Eq. (\ref{asympt})
conserves the momentum, unlike the underlying GP equation (\ref{GP}). This
difference can be explained as it was done above, see Eqs. (\ref{Pgap}) - 
(\ref{dP/dtau}).

In the case of $m_{\mathrm{eff}}\sigma <0$, Eq. (\ref{asympt}) has soliton
solutions. Their exact form is 
\begin{equation}
\Phi (x)=A\exp \left[ i\left( kx-\Delta E\cdot t\right) \right] \,\mathrm{
sech}\left( A\sqrt{g\left\vert m_{\mathrm{eff}}\right\vert }\left(
x-vt\right) \right)  \label{soliton}
\end{equation}
with the wavenumber $k$, velocity 
\begin{equation}
v=k/m_{\mathrm{eff}},  \label{dynamical_mass}
\end{equation}
an arbitrary amplitude $A$, and 
\begin{equation}
\Delta E=-\frac{1}{2}gA^{2}\mathrm{sgn}\left( m_{\mathrm{eff}}\right) +\frac{
1}{2}m_{\mathrm{eff}}v^{2}.  \label{DeltaE}
\end{equation}

Close to the BZ center, i.e., for small $k$, the Bloch function may be
reasonably approximated by a combination of three relevant harmonics, 
\begin{equation}
F(x)=c_{1}\exp (ikx)+c_{2}\exp \left( i(k+2\pi )x\right) +c_{3}\exp \left(
i(k-2\pi )x\right) .  \label{k=0}
\end{equation}
Substitution of this into Eq.~(\ref{F}) yields, in the first approximation, 
$E(k)=E_{0}^{\mathrm{(center)}}+k^{2}/\left( 2m_{\mathrm{eff}}^{\mathrm{\
(center)}}\right) $, with 
\begin{equation}
E_{0}^{\mathrm{(center)}}=\pi ^{2}-\sqrt{\pi ^{4}+\varepsilon ^{2}/2},
\label{E0center}
\end{equation}
and an expression for the effective mass which is indeed positive, 
\begin{equation}
m_{\mathrm{eff}}^{\mathrm{(center)}}=\frac{2\pi ^{4}+\varepsilon ^{2}+\pi
^{2}\sqrt{4\pi ^{4}+2\varepsilon ^{2}}}{10\pi ^{4}+\varepsilon ^{2}-3\pi ^{2}
\sqrt{4\pi ^{4}+2\varepsilon ^{2}}}~.  \label{eff(k=0)}
\end{equation}
Other coefficients in Eqs. (\ref{k=0}) and (\ref{DeltaE}) are found to be 
$c_{1}=1,c_{2}=c_{3}=-E_{0}/\varepsilon $, and 
$g=(1+12c_{2}^{2}+6c_{2}^{4})/(1+2c_{2}^{2})$. Then, with regard to Eqs. (\ref
{soliton}) and (\ref{phi}), the full lowest-order approximation
(corresponding to $k=0$) for the localized state in the attractive model is 
\begin{eqnarray}
\phi (x,t) &=&A\exp \left[ im_{\mathrm{eff}}^{\mathrm{(center)}}vx-i\left(
E_{0}^{\mathrm{(center)}}+\Delta E\right) t\right]  \nonumber \\
&&\times \left[ 1+2c_{2}\cos (2\pi x)\right] \,\mathrm{sech}\left( A\sqrt{
g\left\vert m_{\mathrm{eff}}^{\mathrm{(center)}}\right\vert }\left(
x-vt\right) \right) ,  \label{center}
\end{eqnarray}
where $\Delta E$ is given by Eq. (\ref{DeltaE}). We stress that this soliton
exists only in the attractive model, with $\sigma =-1$.

In a similar way, one can work out an approximation for the linear Bloch
function at the BZ edge, i.e., near $k=\pi $, as a combination of two
harmonics: 
\begin{equation}
F(x)=c_{1}\exp (ikx)+c_{2}\exp \left( i(k-2\pi )x\right) ,  \label{Fedge}
\end{equation}
cf. Eq. (\ref{slowfast}). The substitution of this into Eq.~(\ref{F}) yields 
$E(k)=E_{0}^{\mathrm{(edge)}}+\left( k-\pi \right) ^{2}/(2m_{\mathrm{eff}}^{
\mathrm{(edge)}})$, where 
\begin{equation}
E_{0}^{\mathrm{(edge)}}=(\pi ^{2}\pm |\varepsilon |)/2,  \label{E0edge}
\end{equation}
\begin{equation}
m_{\mathrm{eff}}^{\mathrm{(edge)}}=\frac{|\varepsilon |}{|\varepsilon |\pm
2\pi ^{2}},  \label{eff(k=pi)}
\end{equation}
and $g=3/4$. The signs $+$ and $-$ corresponds to the second and first Bloch
bands, respectively. As is seen, the effective mass (\ref{eff(k=pi)}) for
the linear excitations near the BZ edge for the first Bloch band is negative
if $|\varepsilon |~<2\pi ^{2}$ (below, we present typical numerical results
for $\varepsilon =5$, which satisfies this condition), and the effective
mass for the second Bloch band is positive. These two possibilities
correlate to the conclusion formulated above within the framework of the
approximation based on Eqs. (\ref{slowfast}) and (\ref{gapmodel}), that
there may exist gap solitons with positive and negative masses [that
approximation definitely corresponds to the case of the weak OL, i.e., 
$|\varepsilon |<2\pi ^{2}$, in terms of Eq. (\ref{eff(k=pi)})].

The full form of lowest-order approximation (corresponding to $k=\pi $) for
the soliton carried by the Bloch wave near the BZ edge in the first Bloch
band (the one with the \emph{negative} effective mass) follows from Eq. (\ref
{soliton}) [cf. Eq. (\ref{center})]: 
\begin{eqnarray}
\phi (x) &=&A\exp \left[ im_{\mathrm{eff}}^{\mathrm{(edge)}}vx-i\left(
E_{0}^{\mathrm{(edge)}}+\Delta E\right) t\right]  \nonumber \\
&&\times \cos (\pi x)\,\mathrm{sech}\left( A\sqrt{g\left\vert m_{\mathrm{eff}
}^{\mathrm{(edge)}}\right\vert }\left( x-vt\right) \right) .  \label{edge}
\end{eqnarray}
As it follows from Eq. (\ref{asympt}), this soliton exists in the repulsive
model ($\sigma =+1$) if $|\varepsilon |<2\pi ^{2}$. On the other hand, the
soliton carried by the Bloch wave near the BZ edge in the second band exists
in the attractive model, since it has $m_{\mathrm{eff}}^{(\mathrm{edge})}>0$.

\section{Standing and moving solitons}

\subsection{Zero-velocity solitons}

Stationary soliton solutions can be easily found in the form of $\phi
(x,t)=\phi (x)\exp \left( -iEt\right) $, applying the numerical shooting
method to the stationary version of the underlying GP equation (\ref{GP}).
Figures 1(a) and 1(b) display, respectively, typical examples of the
localized solution for the attractive ($\sigma =-1$) and repulsive ($\sigma
=+1)$ model with the OL potential ($\varepsilon =5$). In the former case,
the analytical approximation (\ref{E0center}) with $\varepsilon =5$ yields
the value $E_{0}=-1.194$ of the chemical potential at the BZ center, and in
the latter case, Eq. (\ref{E0edge}) with $\varepsilon =5$ yields 
$E_{0}=-7.\,\allowbreak 565$ at the BZ edge. 
\begin{figure}[tbh]
\begin{center}
\includegraphics[width=11cm]{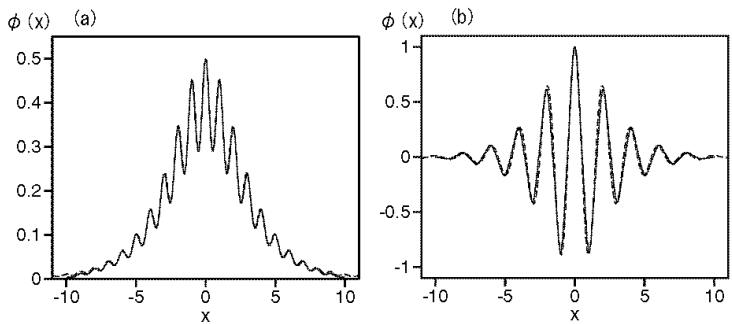}
\end{center}
\caption{Typical examples of numerically found and analytically predicted
(solid and dashed curves, respectively) stable zero-velocity solitons in Eq.
(\protect\ref{GP}) with attraction, $\protect\sigma =-1$ (a), and
repulsione, $\protect\sigma =+1$ (b). The amplitude of the optical-lattice
potential is $\protect\varepsilon =5$.}
\label{fig:1}
\end{figure}

In the example shown in Fig. 1(a), the soliton's amplitude is $A=0.5$ and
its norm is $N=0.354$. In Fig. 1(b), the amplitude is $A=1$, and the norm is 
$N\equiv 0.873$. In both parts of Fig. 1, the dashed profiles show the
analytical predictions (\ref{center}) and (\ref{edge}), respectively,
corresponding to the same values of the amplitude. The comparison with the
analytical approximations makes it obvious that they are very accurate in
both cases. At other values of the parameters, the agreement between the
numerically found and analytically predicted shapes of the solitons is,
generally, as good as in these examples.

It is necessary to stress that, alongside broad solitons extending to many
periods of the OL, a generic example of which is shown in Fig. 1(a), stable
narrow solitons, which are confined, essentially, to a single OL cell, also
exist in the attractive model. They were found in Ref. \cite{Wang}. Quite
similarly, the 2D and 3D counterparts of the 1D equation (\ref{GP}) with
attraction also support both narrow (\textquotedblleft single-cell") and
broad (\textquotedblleft multi-cell") stable solitons \cite{BMS}. 
\begin{figure}[tbh]
\begin{center}
\includegraphics[width=11cm]{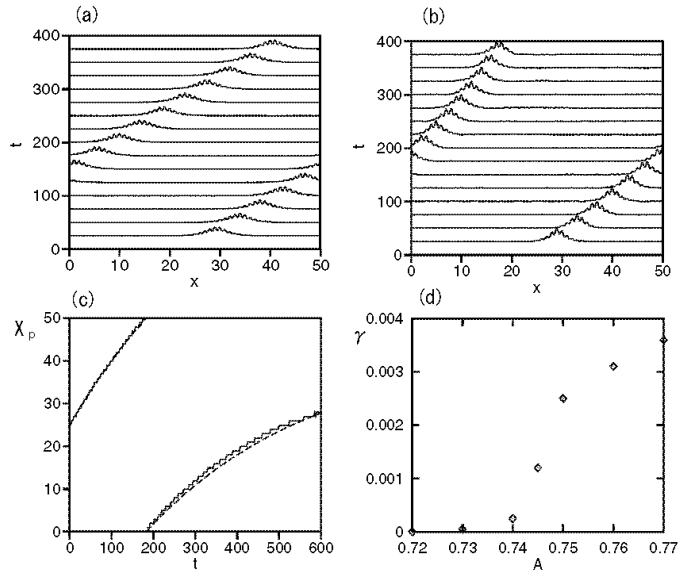}
\end{center}
\caption{Results of numerical simulations of the moving soliton generated by
the initial condition (\protect\ref{phi0attr}) with $k=0.2$. The panels (a)
and (b) here, as well as in Figs. 4 and 5 below, and Figs. 6 and 8 show the
time evolution of $\left\vert \protect\phi (x)\right\vert $. In the panel
(a), the initial amplitude of the soliton is $A=0.5$, while panels (b) and
(c) pertain to $A=0.75$. The panel (c) displays a relatively early stage of
the motion of the soliton's center in the case of braking, the dashed curve
being the fit to Eq. (\protect\ref{v(t)}). The panel (d) shows the braking
rate, defined as in Eq. (\protect\ref{v(t)}), vs. the soliton's amplitude.}
\label{fig:2}
\end{figure}

\subsection{Moving solitons in the attractive model}

The next step is to simulate moving solitons that were also predicted above.
Starting with the attractive model, in Fig. 2 we display a numerically found
moving soliton, which is generated, in the attractive model, by the
analytical waveform (\ref{center}) with $\varepsilon =5$, 
\begin{equation}
\phi _{0}(x)=A~e^{ik(x-L/2)}\left[ 0.877+0.216\cos \left( 2\pi
(x-L/2)\right) \right] \,\mathrm{sech}\left( A(x-L/2)\right) ,
\label{phi0attr}
\end{equation}
taken as the initial condition for Eq.~(\ref{GP}). In this case, periodic
boundary condition were used with the spatial period $L=50$ (results of
simulations with the parabolic trapping potential will be presented below).
The initial wavenumber in Eq. (\ref{phi0attr}) is $k=0.2$; the amplitude is 
$A=0.5$ in the case shown in Fig. 2(a), and $A=0.75$ in Fig. 2(b) (the value
of $k=0.2$ is formally incompatible with the system's period, $L=50$, but
this mismatch plays no practical role, as the soliton's size, $l_{\mathrm{sol
}}$ $\sim A^{-1}$, inside which the phase field carrying the wavenumber was
actually created, is much smaller than $L$).

For $A=0.5$, a stable soliton steadily moving at the velocity $v=0.175$ was
found in the simulations, see Fig. 2(a) (the velocity was identified as that
of the soliton's peak). Following Eq. (\ref{dynamical_mass}), the effective
mass corresponding to it was calculated as $k/v$, with the wavenumber $k$ of
the initial configuration (\ref{phi0attr}). The thus found effective mass 
$1.143$ is very close to the analytical value $1.13$ produced by Eq. (\ref
{eff(k=0)}).

For $A=0.75$, the simulations produce an altogether different result: the
soliton's velocity slowly decreases with time, see Figs. 2(b) and 2(c). At a
relatively early stage of the evolution, the time dependence of the velocity
may be well fitted by 
\begin{equation}
v(t)=v_{0}\exp (-\gamma t),\,\mathrm{~ with}\,v_{0}=0.17,\gamma =0.0025.
\label{v(t)}
\end{equation}
To illustrate this, the law of motion $X_{p}(t)=25+0.17(\left[ 1-\exp
(-0.0025t)\right] /0.0025$, which is the temporal integration of Eq. (\ref
{v(t)}), is depicted in Fig.~2(c) by a dashed curve, and the actual
trajectory of the soliton's center is shown by a continuous line.

Figure 2(d) displays the braking rate $\gamma $, which is defined as in Eq. 
(\ref{v(t)}), i.e., by fitting the initial law of motion to the exponential
decay of the velocity, as a function of the soliton's amplitude $A$. This
dependence was obtained from systematic simulations of Eq. (\ref{GP}) with
the fixed OL strength, $\varepsilon =5$, and different initial conditions.
The braking sets in at the critical value of the amplitude, $A_{\mathrm{cr}
}=0.72$. It is relevant to mention that a qualitatively similar effect was
observed, in various forms, in simulations of the motion of a \emph{discrete}
soliton in the NLS lattice equation: if the amplitude of the moving soliton
exceeds a critical value, it quickly gets trapped by the lattice 
\cite{Dima}, otherwise it moves freely \cite{Feddersen}.

\begin{figure}[tbh]
\begin{center}
\includegraphics[width=6cm]{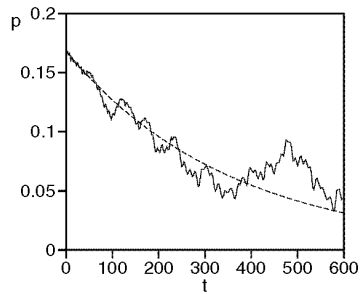}
\end{center}
\caption{The decay of the soliton's momentum, in the same case as shown in
Figs. 2(b) and 2(c), on a longer time scale. The dashed curve is the fit to
Eq. (\protect\ref{v(t)}).}
\label{fig:3}
\end{figure}

At a later stage of the evolution, the soliton's motion law strongly
deviates from the exponential relaxation assumed in Eq. (\ref{v(t)}). To
illustrate this, in Fig. 3 we display the evolution of the soliton's
momentum per atom, which is computed from the numerical data as 
$p=i\int_{-\infty }^{+\infty }\left( \partial \phi ^{\ast }/\partial x\right)
\phi (x)dx/\int_{-\infty }^{\infty }|\phi (x)|^{2}dx$ [cf. Eq. (\ref{Pphi})], 
at a moderately long time scale. The initial exponential decay, 
$p(t)\approx 0.168\exp (-0.0028t)$, with the decay rate very close to that in
Eq. (\ref{v(t)}), changes into an erratic motion at a later stage, with the
deceleration sometimes changed by acceleration. This picture suggests that
dynamical chaos may develop in the motion of the soliton. To further
investigate the issue, in Fig. 4 we display continuation of the motion to
extremely long time. The figure, and, especially, the \emph{continuous}
power spectrum in panel (c) clearly demonstrate that chaotic motion sets in
indeed.  However, the soliton's shape remains strictly coherent at all times,
hence that the mean-field approximation based on the GP equation remains
relevant in this case, unlike situations when the wave field \emph{as a whole}
becomes chaotic, and the BEC disintegrates, making the GP equation
irrelevant \cite{chaos}. 
\begin{figure}[tbh]
\begin{center}
\includegraphics[width=13cm]{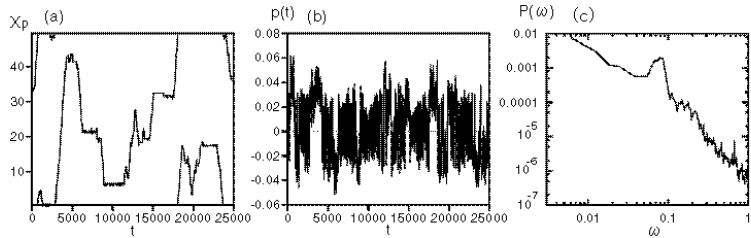}
\end{center}
\caption{The continuation of the Fig. 3 to an extremely long time scale. The
panels (a) and (b) display the time dependences of the coordinate of the
soliton's center and momentum per atom, and the panel (c) is the power
spectrum of the latter in the double-logarithmic scale.}
\label{fig:4}
\end{figure}

\subsection{Relaxation of a two-soliton in the attractive model}

It is well known that the usual self-focusing NLS equation, i.e., Eq. (\ref
{GP}) with $\sigma =-1$ and $\varepsilon =0$, gives rise not only to
fundamental solitons, but also to their higher-order counterparts, the
simplest one being the two-soliton (a breather), which is generated by the
initial condition with the double amplitude. We simulated evolution of
similar configurations in the presence of the OL potential, taking, for
example, the initial condition as in Eq. (\ref{phi0attr}) but with the
amplitude $2A$ instead of $A$ [while $\mathrm{sech}\left( A(x-L/2)\right) $
was not altered]. If the amplitude is smaller than some critical value, this
configuration indeed evolves into a persistent breather, see an example in
Fig. 5(a) for $A=0.16$. However, a \textquotedblleft heavier" initial
two-soliton configuration relaxes into a stationary fundamental soliton,
like in Fig. 5(b) in the case of $A=0.3$. Figure 5(c) additionally displays
the evolution of the pulse's amplitude in the latter case.
\begin{figure}[tbh]
\begin{center}
\includegraphics[width=15cm]{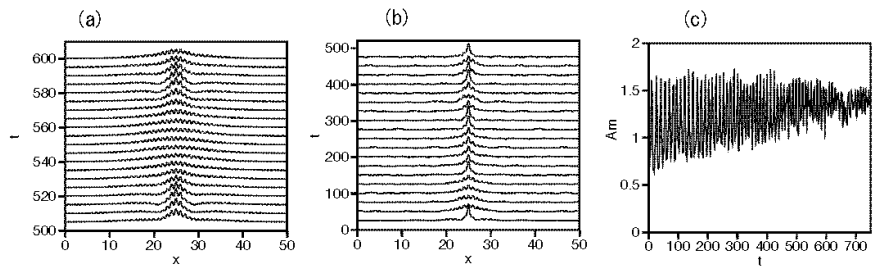}
\end{center}
\caption{Evolution of the two-soliton, initiated by the initial
configuration (\protect\ref{phi0attr}) with the double amplitude: (a) $A=0.16
$; (b) and (c): $A=0.3$. A persistent breather appears in the former case,
while in the latter case the pulse relaxes (without conspicuous radiation
loss) into a fundamental soliton.}
\end{figure}

It is relevant to mention that the attractive model with the OL potential
may have stable static two-soliton (or multi-soliton) solutions of a
different type, when two narrow solitons are trapped in two adjacent wells
of the OL potential. Such states and their stability limits were found in
Ref. \cite{Wang} in terms of a nonlinear-optical model.

\subsection{Moving solitons in the repulsive model}

Motion of solitons in the repulsive GP equation (\ref{GP}), with $\sigma =+1$
, was systematically investigated too. Again fixing $\varepsilon =5$, we
take, as a generic example, the initial condition in the form suggested by
the analytical approximation (\ref{edge}), i.e.,
\begin{equation}
\phi _{0}(x)=1.985~A~e^{ik(x-L/2)}\cos \left( \pi (x-L/2)\right) \mathrm{sech
}\left( A(x-L/2)\right) ,  \label{phi0rep}
\end{equation}
cf. Eq. (\ref{phi0attr}). The outcome of the simulation, corresponding to 
$k=0.1$ in Eq. (\ref{phi0rep}), is shown in Fig. 6, which displays the
evolution of $\left\vert \phi (x)\right\vert $ for two different amplitudes,
(a) $A=0.30$ and (b) $A=0.55$. In the former case, the gap soliton moves
steadily at the velocity $v=-0.274$. The effective mass corresponding to
this soliton, as determined from the numerical data according to the same
definition as above, $m_{\mathrm{eff}}=k/v$, is $-0.365$, which is close
enough to the analytical value $-0.34$ produced by Eq. (\ref{eff(k=pi)}).
\begin{figure}[tbh]
\begin{center}
\includegraphics[width=14cm]{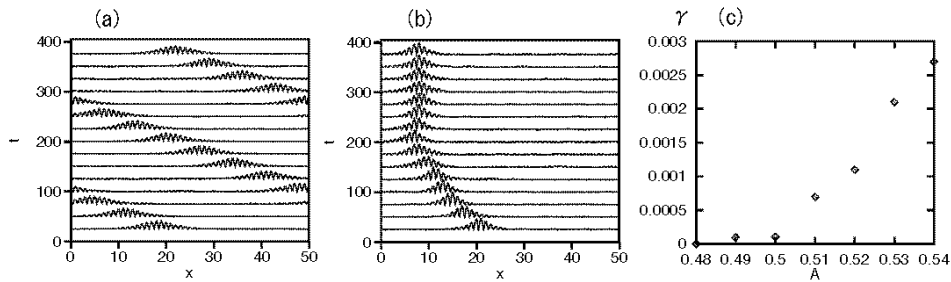}
\end{center}
\caption{The panels (a), (b) and (c) show the same as (a), (b) and (d) in
Fig. 2, but for the case when the soliton is launched in the repulsive
model, using the initial condition (\protect\ref{phi0rep}) with $k=0.1$. The
panels (a) and (b) correspond to the amplitudes $A=0.30$ and $0.55$,
respectively.}
\end{figure}

Figure~6(b) shows that a heavier soliton, with $A=0.55$, does not move
steadily. Instead, it starts to brake, like heavy solitons in the attractive
model, see Fig. 2. The velocity decays exponentially at the initial stage of
the evolution, like in Eq. (\ref{v(t)}), the corresponding damping rate 
$\gamma $ being displayed in Fig.~6(c) versus $A$. As well as in the
attractive model, the transition from the steady motion to the braking
occurs at a critical value of the soliton's amplitude, which is $A_{\mathrm{
\ cr}}\approx 0.48$ in this case.

\section{Solitons in the parabolic trap or anti-trap}

Any experimental setup in BEC uses a parabolic trap \cite{Pit} (bright 1D
solitons were observed in an expulsive potential created by an inverted trap 
\cite{soliton_exper}). This means that the underlying GP\ equation 
(\ref{GP}) should be extended to the form 
\begin{equation}
i\frac{\partial \phi }{\partial t}=-\frac{1}{2}\frac{\partial ^{2}\phi }{
\partial x^{2}}+\left[ \sigma |\phi |^{2}-\varepsilon \cos \left( 2\pi
x\right) +B\left( x-\frac{L}{2}\right) ^{2}\right] \phi ,  \label{GPtrap}
\end{equation}
where $B$ is the strength of the trap, and its center is set at $x=L/2$.

\begin{figure}[tbh]
\begin{center}
\includegraphics[width=13cm]{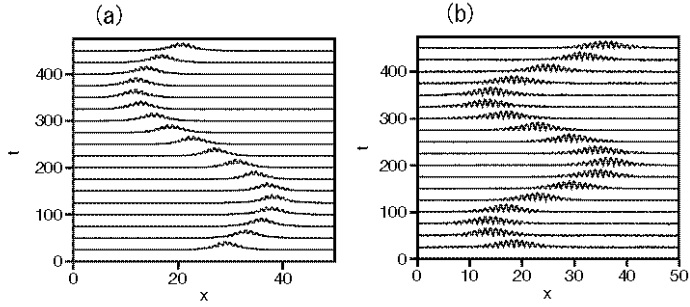}
\end{center}
\caption{Motion of the soliton in the weak parabolic potential: (a) the trap
with $B=0.0001$; (b) the inverted potential (\textit{anti-trap}) with 
$B=-0.0001$. The strength of the optical lattice is $\protect\varepsilon =5$.
In the cases shown in (a) and (b), the solitons were set in motion by
lending them the initial velocities $v_{0}=0.172$ and $-0.270$, respectively.
}
\label{fig:7}
\end{figure}
Figure 7(a) shows the evolution of $\ \left\vert \phi (x)\right\vert $ in
the attractive model ($\sigma =-1$) with $\varepsilon =5$ and a weak trap, 
$B=0.0001$, as obtained from numerical simulations of Eq. (\ref{GPtrap}). The
initial condition is again taken in the form of Eq. (\ref{phi0attr}), which
is suggested by the analytical approximation (\ref{center}), with $A=0.5$
and $k=0.2$. In this case, the soliton performs harmonic oscillations in the
trap, and the motion of the soliton's center-of-mass coordinate, which we
define as 
\begin{equation}
X_{p}(t)\equiv \int_{-\infty }^{\infty }x|\Phi |^{2}dx/\int_{-\infty
}^{\infty }|\Phi |^{2}dx,  \label{Xp}
\end{equation}
is well approximated by the expression 
\begin{equation}
X_{p}(t)=25+13\sin (0.0132~t).  \label{oscillations}
\end{equation}

To proceed with an analytical approach to the soliton's motion in the
parabolic potential, we insert the same ansatz (\ref{phi}), which was used
above, into Eq. (\ref{GPtrap}), and derive a modified version of Eq. (\ref
{asympt}), 
\begin{equation}
i\Phi _{t}=-\left( 1/2\right) m_{\mathrm{eff}}\Phi _{xx}+\left[ \sigma
g|\Phi |^{2}+B(x-L/2)^{2}\right] \Phi .  \label{NLStrap}
\end{equation}
Then, the following relation is an exact corollary of Eq. (\ref{NLStrap}), 
\begin{equation}
m_{\mathrm{eff}}\ddot{X}_{p}=-2B(X_{p}-L/2),  \label{osc}
\end{equation}
where the overdot stands for the time derivative. In the model with
attraction, where the effective mass is positive, Eq. (\ref{osc}) predicts
harmonic oscillations of the soliton with the frequency $\omega _{\mathrm{ho}
}=\sqrt{2B/m_{\mathrm{eff}}}$. For instance, in the example shown in Fig.
7(a), Eq. (\ref{eff(k=0)}) with $\varepsilon =5$ yields $m_{\mathrm{eff}
}=1.13$, hence, with $B=0.0001$, we predict the frequency $\omega _{\mathrm{
ho}}=0.0133$, which is to be compared with the numerically found value 
$\omega _{\mathrm{\ num}}=0.0132$ in Eq. (\ref{oscillations}).

Figure 7(b) shows the evolution of $|\phi (x)|$ in the repulsive model with
the \emph{inverted} parabolic potential, $B=-0.0001$. The initial condition
is again taken in the form of Eq. (\ref{phi0rep}), with $A=0.3$ and $k=0.1$.
It is clearly observed that the negative-mass soliton is \emph{trapped} by
the \emph{anti-trapping} potential. The motion of the soliton's center is
fitted by 
\begin{equation}
X_{p}(t)=25-11.5\sin (0.0235~t),  \label{anti-oscillations}
\end{equation}
cf. Eq. (\ref{oscillations}). In this case, the analytical approximation 
(\ref{eff(k=pi)}) with $\varepsilon =5$ yields $m_{\mathrm{eff}}=-0.339$,
hence the corresponding harmonic-oscillation frequency with $B=-0.0001$ is 
$\omega _{\mathrm{ho}}=\sqrt{2B/m_{\mathrm{eff}}}=\allowbreak 0.0243$, which
is close to the numerically found value $\omega _{\mathrm{num}}$ that
appears in Eq. (\ref{anti-oscillations}).

We have checked too that, in the case of the usual trap ($B>0$), the
negative-mass soliton in the repulsive model is indeed expelled from the
trap region, as it should be expected. In that case, its observed motion
again complies well with Eq. (\ref{osc}).

\begin{figure}[tbh]
\begin{center}
\includegraphics[width=11cm]{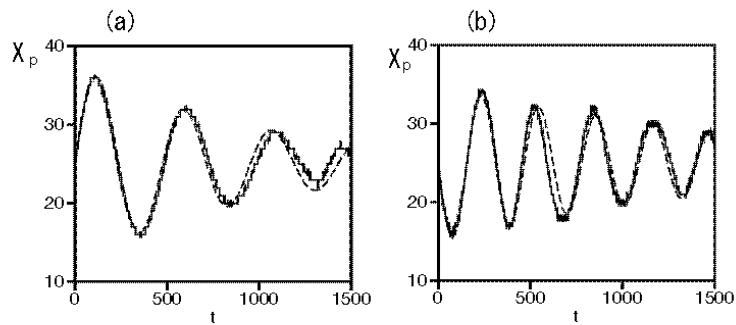}
\end{center}
\caption{Numerically found trajectories of the soliton's center in the
presence of the harmonic trap or anti-trap, in the case when the soliton's
amplitude exceeds the braking threshold. The GP equation (\protect\ref
{GPtrap}) was simulated, with $\protect\sigma =-1$ and $B=0.001$ (attraction
and normal trap) in (a), and with $\protect\sigma =+1$ and $B=-0.001$
(repulsion and anti-trap) in (b). The dashed curves show the respective
analytical fits (\protect\ref{damped_attr}) and (\protect\ref{damped_rep}).}
\label{fig:8}
\end{figure}

Lastly, we have also found that, as well as in the absence of the parabolic
potential (see above), the motion of the soliton is subject to braking if
its amplitude is too large. In that case, the positive- and negative-mass
solitons perform damped oscillations in the trapping or anti-trapping
potential, respectively. Typical examples are shown in Fig. 8, which
displays the motion of the soliton's center as found from the simulations
initiated by the same initial conditions (\ref{phi0attr}) and (\ref{phi0rep}) 
as above, but with larger amplitudes: $A=0.75$ in (a), and $A=0.52$ in
(b). In these two cases, the damped oscillations are well approximated by
analytical expressions 
\begin{equation}
X_{p}(t)=25+12.5\sin (0.0132t)\exp (-0.0010~t)  \label{damped_attr}
\end{equation}
[the dashed curve in Fig. 8 (a)], and 
\begin{equation}
X_{p}(t)=25-10\sin (0.0202~t)\exp (-0.0006~t)  \label{damped_rep}
\end{equation}
[the dashed curve in Fig. 8 (b)].

At other values of the parameters, the motion of the positive- and
negative-mass solitons in the trapping and anti-trapping potentials is the
same as shown in the above examples. In all the cases when the braking did
not set in, the numerically simulated motion (both oscillations and
expulsion, depending on the signs of the effective mass and the potential)
was found to be in good agreement with the predictions based on Eq. (\ref
{osc}).

The trapping of the gap soliton by the \emph{inverted} parabolic potential
in the \emph{repulsive} BEC\ loaded into an OL is a challenge for
experimental verification. We stress that, while the gap soliton should be
trapped by the inverted potential, this is definitely not expected for a
plain broad distribution of the atomic density, in which case the
anti-trapping potential will try to expel atoms in the usual way (of course,
this trend may be countered by trapping the atoms in the OL). Thus, the
trapping in the inverted parabolic potential is a specifically nonlinear
dynamical effect, endemic to the negative-mass solitons.

\section{Collisions between oscillating solitons}

If the direct or inverted parabolic potential retains the solitons, it is
quite easy to create an initial configuration with two well-separated
identical solitons, and lend them opposite initial velocities (or just zero
velocities, see below). Then, they will start to oscillate in opposite
directions, and will thus have to collide periodically. This setting is
quite convenient to test the character of collisions between solitons.

\begin{figure}[tbh]
\begin{center}
\includegraphics[width=13cm]{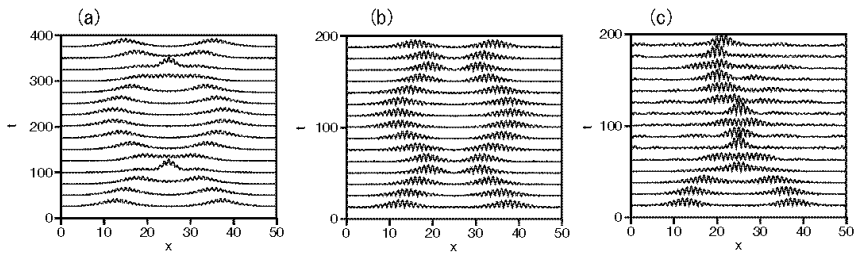}
\end{center}
\caption{Collisions between positive-mass (a) and negative-mass (b,c)
solitons which oscillate, respectively, in the trapping or anti-trapping
parabolic potential with the strength $B=\pm 0.0001$. The difference between
the cases shown in (b) and (c) is in the initial phase difference $\Delta 
\protect\varphi _{0}$ between the negative-mass solitons: $\Delta
\varphi _{0}=\protect\pi $ in (b), and $\Delta \protect\varphi _{0}=0$ in
(c). }
\label{fig:9}
\end{figure}
We simulated this situation in the attractive model, creating an initial
configuration $\phi _{0}(x)$ which was a linear superposition of the
analytical approximations (\ref{center}) for two solitons. As a typical
example, in Fig. 9(a) we display the case when centers of the two solitons
with equal amplitudes $A=0.3$ are set at the points $x_{1}=L/2-13$ and 
$x_{2}=L/2+13$, so that 
\begin{eqnarray}
\phi _{0}(x) &=&A\left\{ \left[ 0.877+0.216\cos 2\pi (x-x_{1})\right] 
\mathrm{sech}\left( A(x-x_{1})\right) \right.   \nonumber \\
&&+\left. \left[ 0.877+0.216\cos 2\pi (x-x_{2})\right] \mathrm{sech}\left(
A(x-x_{2})\right) \right\} .  \label{phi0coll_attr}
\end{eqnarray}
Note that the initial wavenumbers in this expression are set equal to zero
[cf. Eq. (\ref{phi0attr})], hence the initial velocities of the two solitons
are zero too; however, being in-phase, the solitons attract each other,
which initiates their oscillatory motion. Figure 9(a) clearly demonstrates
(which is also corroborated by extending the simulations to much longer
times) that the positive-mass solitons collide repeatedly, each time passing
through each other, and remain completely stable.

A typical example of a gap-soliton pair that survives indefinitely many
collisions in the repulsive BEC placed in the inverted parabolic potential
is displayed in Fig. 9(b). The two solitons were initially taken as a linear
superposition of two analytical waveforms (\ref{edge}), again with equal
amplitudes $A=0.3$, centers placed at the points $x_{1}=L/2-13$ and 
$x_{2}=L/2+13$, and zero initial wavenumbers: 
\begin{equation}
\phi _{0}(x)=1.985A\left[ \cos \left( \pi (x-x_{1})\right) \mathrm{sech}
\left( A(x-x_{1})\right) -\cos \left( \pi (x-x_{2})\right) \mathrm{sech}
\left( A(x-x_{2})\right) \right] .  \label{phi0coll_rep}
\end{equation}

Unlike the initial condition (\ref{phi0attr}) which was used in the
attractive model, the initial phase difference between the solitons in the
configuration (\ref{phi0coll_attr}) is $\Delta \varphi _{0}=\pi $, hence
they repel each other. In this case too, the solitons perform indefinitely
many oscillations, colliding repeatedly without developing any instability.
Note that, in contrast with the case shown in Fig. 9(a), this time the
colliding solitons do not pass through each other, but rather bounce, which
is explained by the fact that they keep the phase difference $\pi $.

However, the solitons do not always collide elastically. Another generic
outcome is \emph{fusion} of two in-phase solitons into a single one. For
example, in the repulsive model with the anti-trapping parabolic potential
and the same initial configuration as in Eq. (\ref{phi0coll_rep}), but this
time with $\Delta \varphi _{0}=0$, 
\[
\phi _{0}(x)=1.985A\left[ \cos \left( \pi (x-x_{1})\right) \mathrm{sech}
\left( A(x-x_{1})\right) +\cos \left( \pi (x-x_{2})\right) \mathrm{sech}
\left( A(x-x_{2})\right) \right] ,
\]
we have found that the first collision leads to merger of the solitons into
one pulse, which is accompanied by some radiation loss, see Fig. 9(c). The
resultant single soliton then performs harmonic oscillations with a small
amplitude. Qualitatively, the fusion may be explained the following way: in
the course of the collision, overlapping between the in-phase solitons leads
to increase of the amplitude past the threshold at which the braking sets in
(see above).

Further simulations demonstrate that, in the case shown in Figs. 9(b) and
9(c), the colliding solitons eventually merge into one if the initial phase
difference between them belongs to the interval $\left\vert \Delta \varphi
_{0}\right\vert \leq 0.8\pi $. It is noteworthy that, when $\Delta \varphi
_{0}$ is close to the critical value ($0.8\pi $, in this example), the
fusion happens after several collisions.

In the attractive model, the merger induced by the collisions was observed
too, if the solitons' amplitude was large enough. On the other hand, in both
models (repulsive and attractive) merger of solitons with $\Delta \varphi
_{0}=\pi $ has not been observed. We stress that (as well as in the case of
the discrete NLS equation \cite{Dima}) solitons with very large amplitudes
cannot collide at all, as they will be quickly halted by the braking force
induced by the OL.

\section{Conclusion}

In this work, we aimed to investigate, in a systematic way, motion of 1D
solitons in the attractive and repulsive Bose-Einstein condensates (BECs)
loaded into an optical lattice (OL) and subject to the action of an external
parabolic potential. First of all, we have demonstrated, in two different
analytical forms, that, in the repulsive BEC, when the soliton is of the gap
type, its effective mass is negative. This gives rise to the prediction that
calls for experimental verification: in the latter case, the soliton cannot
be confined by the usual trapping parabolic potential, but it can be held by
the \emph{anti-trapping} inverted potential. We have also investigated the
motion of the solitons in long systems, concluding that, in both cases of
the positive and negative mass, the soliton moves unhindered if its
amplitude is below a certain threshold; above the threshold, the soliton is
braked by its interaction with the OL. At a late stage of the braking, the
damped motion becomes chaotic.

We have also investigated the evolution of the two-soliton state in the
attractive model. It was found that the two-soliton generates a persistent
breather, if its amplitude is not too large; otherwise, the two-soliton
suffers fusion into a single fundamental soliton.

Lastly, we have considered collisions of two solitons trapped in the
parabolic potential. Depending on their amplitudes and phase difference $
\Delta \varphi $, the solitons perform stable oscillations, colliding
indefinitely many times, or they merge into a single soliton, after one or
several collisions. However, the merger does not occur in the case of $
\Delta \varphi =\pi $.

As concerns values of physical parameters at which the phenomena predicted
in this work can be observed in the experiment, they are not different from
those for which other soliton effects in OLs were recently predicted \cite
{othergapsol,otherOLsol} (the number of atoms in the soliton may be in the
range of $10^{3}-10^{4}$, the number of the OL periods inside the parabolic
trap or anti-trap can easily be $\sim 100$ or larger, etc.). Thus, the
observation of the gap-soliton's confinement in the anti-trapping potential
and other dynamical phenomena predicted in this work seems quite feasible.
Besides that, all the predictions may also be translated into those for
solitons in nonlinear photonic crystals. In particular, the capture of the
gap soliton by the inverted potential implies that a stable bright optical
soliton is possible in a self-defocusing nonlinear anti-waveguide, with the
refractive index periodically modulated in the transverse direction.

Results reported in this paper can be extended to the two-dimensional (2D) case.
In particular, we have demonstrated that a 2D gap soliton in a repulsive BEC
loaded in a 2D optical lattice may be trapped by an 
{\em inverted} isotropic harmonic potential. In that case, the trapped soliton 
demonstrates various modes of stable motion, including oscillations with
periodic passage through the center, and circular motion around the
center. These results will be reported in detail elsewhere.

\section*{Acknowledgement}

B.A.M. appreciates hospitality of the Optoelectronic Research Centre at the
City University of Hong Kong and discussions with W.C.K. Mak. The work of
this author was supported, in a part, by the Israel Science Foundation
through the grant No. 8006/03.

\end{document}